# Thermal Diffusion and Quench Propagation in YBCO Pancake Coils

## Wound with ZnO-and Mylar Insulations

M.D. Sumption[1], M. Majoros[1], M. Susner[1], D. Lyons[2], X. Peng[2], C.F. Clark[3],

W.N. Lawless[3,4], and E.W. Collings[1]

[1]MSE, The Ohio State University, Columbus, OH 43210

[2]Hyper Tech, Columbus, OH 43201

[3]CeramPhysics, Columbus OH 43081

**Abstract**

The thermal diffusion properties of several different kinds of YBCO insulations and the quench properties of pancake coils made using these insulations were studied. Insulations investigated include Nomex, Kapton, and Mylar, as well as insulations based on ZnO, $Zn_2GeO_4$, and ZnO-Cu. Nomex, Kapton, and Mylar, chosen for their availability and ease of use, were obtained as thin ribbons, while the ZnO based insulations were chosen for their high thermal conductivity and were applied by a thin film technique. Initially, short stacks of YBCO conductors with interlayer insulation, epoxy, and a central heater strip were made and later measured for thermal conductivity in liquid nitrogen. Subsequently, three different pancake coils were made. The first two were smaller, each using one meter total of YBCO tape present as four turns around a G-10 former. One of these smaller coils used Mylar insulation co-wound with the YBCO tape, the other used YBCO tape onto which ZnO based insulation had been deposited. One larger coil was made which used 12 total meters of ZnO-insulated tape and had 45 turns. Temperature gradients were measured and thermal conductivities were estimated from these coils, the results obtained were compared to those of the short stacks. The results for all short sample and coil thermal conductivities were ~1-3 $Wm^{-1}K^{-1}$. The lack of distinction for the ZnO-based insulations was attributed to the presence of a thermal interface contact resistance. The ZnO insulations, while not strongly increasing the average thermal conductivity of the winding pack or coil, were much thinner than the other insulations, and would thus enable substantial increases in winding pack critical current density. Finally, quench propagation velocity measurements were performed on the coils (77 K, self field) by applying a DC current and then using a heater pulse to initiate a quench. Normal zone propagation velocity (NZP) values were obtained for the coils both in the radial direction and in the azimuthal direction. Radial NZP values (0.05-0.7 mm/s) were two orders of magnitude lower than axial values (~14-17 mm/s). Nevertheless, the quenches were generally seen to propagate radially within the coils, in the sense that any given layer in the coil is driven normal by the layer underneath it. This initially surprising result is due to the fact that while the radial normal zone propagation velocity (NZP) is much lower than the NZP along the conductor (~100 x) the distance the normal zone must expand longitudinally is much larger than what it must expand radially to reach the same point, in our case this ratio is ~ 1600.

*Footnote 4: Late of Ceram Physics. Bill Lawless passed away Christmas day, 2009.*

## 1.0 Introduction

### 1.1. Background

Most of the quench and overcurrent studies for YBCO coated conductors have been directed towards fault current limiters, or other AC devices, although some initial results related to quenching in strands has been performed [1-5]. Some results on early conductors showed normal zone propagation velocities of 2-5 mm/s at 80 K and 50-98% of $I_c$ [1], measurements of more recent conductors have typically given 10-20 mm/s [2-5]. The measurement processes varied; sometimes they were performed in vacuum, or more frequently under gas cooling. More recently, results for quench measurements of small coils of YBCO are becoming available [6,7], although much work remains to be done. Of course, as it was for other magnet/conductor systems, insulation is very important. One of the key issues is that YBCO coated conductors are very thin, and thus in order to have good thermal transport and high engineering critical current density in the winding (which we denote $J_e$) a thin and thermally conductive insulation is needed. YBCO coils have typically used a thin film insulation, co-wound with the conductor, such as Nomex, Kapton, or Mylar. However, these insulations are somewhat thicker than would be desired, and can limit $J_e$ and thermal conduction. Another alternative is the use of a thin sol-gel type insulation, similar to that used for Bi-2212 [8], bearing in mind that those of Ref [8] tend to stick best to Ag surfaces. But presently, researchers looking at YBCO for magnet applications tend to use Mylar or similar insulation – even if noting the need for better thermal conductivity [9]. Below we will investigate both quench propagation in small pancake coils made with coated conductor and the differences seen with different insulations methods. One of the key drivers for this work has been the need to increase the NZP, in order to enhance magnet protection, this is also investigated below.

*1.2 Outline*

This work focuses on the measurement of thermal gradients and quench propagation in single layer YBCO coils insulated with Mylar and ZnO. There were two motivations for this work. First, we wish to explore quench propagation in YBCO coils. While quenching in BSCCO based coils has been extensively studied, the materials properties and high aspect ratios of coated conductors changes quench propagation in interesting ways. Secondly, we wish to investigate the possibility of speeding the quench of YBCO coils by allowing radial transport of the quench zone.

The idea of radial quenching in YBCO coils was suggested by Oberly [10] as a possible way to mitigate problems related to the slow normal zone propagation (NZP) in YBCO conductors. It is well known that the time required for longitudinal quench propagation along YBCO tape at 77 K is roughly three orders of magnitude greater that that of NbTi or $Nb_3Sn$ at 4.2 K. The normal zone propagation speed for NbTi is some meters per second (m/s), while that of YBCO is some few mm/s to some 10's of mm/s [1-5]. This leads to problems in magnet protection [11]. Under static conditions (i.e., without ramping fields or currents) there is a minimum energy required to cause a quench, the minimum quench energy (MQE). At lower perturbation energies the magnet may recover, but once the MQE is reached, it will proceed to a full quench. At this point, the currents are transferred into the surrounding matrix, thereby heating it and causing the quench to propagate. The speed of the propagation depends on a number of factors, including the heat capacity of the conductor, its resistivity in the normal state, and the level of current in the cross section [11]. As this speed slows, the time required to detect the quench increases, and a smaller region must absorb the energy being dumped into the magnet. Ultimately, the power to the magnet must be shut off as soon as possible after the quench, perhaps using quench heaters to drive larger sections normal (although the MQE needed for YBCO is much higher, and may make this scheme more difficult).

For clarity, when discussing the pancake coil measurements and analysis below, we can consider the pancake coil to lie in the $r$-$\vartheta$ plane, axial to refer to properties along the $z$-axis, azimuthal to refer to properties in the $\vartheta$ direction, and radial to refer to properties in the $r$-direction. The longitudinal direction in the tape (i.e., along the direction of the current flow) then corresponds to the azimuthal direction in the coil, and the tape perpendicular direction corresponds to a radial direction in the coil. The tape parallel direction is associated with properties either along the azimuthal-longitudinal direction, or along the axial direction in the coil. The fact that normal zone velocities vary with direction is well known, harking back to the beginning days of stability studies on low $T_c$ conductors [12-14]. Indeed, many of the essential aspects of strand and magnet stability are very similar to those of low $T_c$ materials [15], and the 3-D nature of quench propagation is well understood in principle [11,15]. However, while truly detailed and 3-D models have developed, usually 1-D models dominated by longitudinal normal zone propagation frame the conceptual approach to thinking about quench in magnets. This has also been the starting point for considering quench in YBCO magnets, with the focus on coil-azimuthal propagation of the quench, corresponding to a longitudinal propagation within a tape. At first glance this makes sense, since the longitudinal propagation in an isolated tape is much faster than the expected radial propagation within a coil -- at least in absolute terms. This scenario implicitly assumes that the insulation between adjacent layers will have a lower thermal diffusion constant, causing quench propagation to occur along the length of the tape, even when it is wound into a coil. However, as is especially true for racetrack magnets (but also for solenoids) the distance around the circumference (or path) of the magnet is much longer than the radial turn-to-turn distance given the very thin YBCO conductor. Thus, it was postulated that if the thermal diffusion in the radial direction could be enhanced, the quench could be made to propagate radially within the coil at the same time as it propagated longitudinally along the wire (azimuthally within the coil), thus causing a much larger fraction of the coil to quench in a given time.

In order to increase the thermal conductivity of the winding pack, we investigated the use of thin ZnO-based insulations known to be electrically insulating and thermally conductive. A series of ZnO based insulations have been developed at CeramPhysics, and were the basis of the insulating materials used here [16,17]. The thermal properties of pure- and varistor-grade ZnO were first investigated by Lawless and Gupta [16] in the temperature range of 1.7-25 K. The thermal conductivities of both of these materials were found to be large (~10 $Wm^{-1}K^{-1}$) at 25 K. As a result, they were then proposed [16] for use as insulation in HTS coils. The goal was to identify materials that not only provided thinner electrical insulation, but also improved quench protection because of their large thermal conductivity. The thermal conductivities of pure- and varistor-grade ZnO were found to be a nearly constant 62 $Wm^{-1}K^{-1}$ in the temperature range 60-90 K. As a result, pure ZnO was chosen as the material of interest for further investigation as an insulator for YBCO. It has several desirable properties in addition to its large thermal conductivity: low cost, and ease of application by sputtering. Of course, because the winding must be epoxy impregnated, the inter-tape regions will contain epoxy as well as insulation, both of which have relatively low thermal conductivity. With Nomex, Kapton, or Mylar, the insulation layer is impermeable to the epoxy. On the other hand, with a ceramic insulation like ZnO, the epoxy may interpenetrate cracks in the relatively thin ZnO layer. Thus in both cases the final thermal properties of the interlayer are some combination of the epoxy and the insulation.

In this work, we compare thermal conductivity and quench propagation in YBCO pancake coils wound with ZnO based insulations to one wound with Mylar insulation. We first measured the thermal conductivities of model winding stacks: simple stacks of conductors instrumented for temperature and voltage. The thermal gradients in ZnO and doped ZnO insulated YBCO stacks were measured and compared to those in stacks insulated with Nomex, Kapton, and Mylar. The average thermal conductivity of the ZnO insulated stacks were not very different than that of the Nomex, Mylar, and Kapton stacks, presumably due to the presence of a thermal interface contact resistance. However, the ZnO insulation was much thinner than the other insulations, such that the winding $J_e$ was significantly improved. Following the stack measurements, several YBCO pancake coils were made, both with ZnO insulation as well as with Mylar insulation. Thermal gradients were measured in these coils and compared to the tape stack measurement results. Finally, NZP measurements were made on the coils, and the results for ZnO-insulated coils were compared to those for Mylar-insulated coils. The ZnO coils were found to have the highest packing density but their thermal conductivities and NZPs were found to be only slightly increased. However, it was found that quenches tended to propagate radially both for the ZnO and also the Mylar insulated coils, due to a large coil perimeter/conductor thickness ratio.

**2.0 Sample Preparation and Experimental Measurements**

Two different IBAD-architecture tapes were obtained from Superpower, Conductor 1 and Conductor 2 (see Table 1). Both conductors 1 and 2 were used for short stack measurements (see below), while all of the coils used in this study were wound with conductor 2. Thee insulations were available as ribbons; Nomex, Kapton, and Mylar, their thicknesses were 38 µm, 40 µm, and 19 µm, respectively. Short stack samples were used for 1-D thermal propagation measurements, while the coils were used for in-situ thermal propagation measurements and NZP measurements. All measurements were performed at 77 K in $LN_2$.

*2.1. Preparation of Short Stacks of YBCO and Thermal Measurements*

In preparation for 1-D thermal propagation measurement, YBCO tape was cut into segments approximately 4 cm in length. These segments were then stacked and epoxy impregnated using Stycast 1266 epoxy and a wet lay-up technique (the tapes were dipped in epoxy and then assembled). Stycast 1266, and thin, and moderately thermally conductive epoxy (as opposed to Stycast blue (2850FT)) was chosen to mimic the CTD 101 used for vacuum impregnation of the coils [18,19], since vacuum impregnation with CTD101 was difficult to achieve for these short samples (the thermocouples were damaged). A pressure of 8.9 MPa was applied during drying to simulate a typical expectation for a magnet winding. A heating strip, which consisted of a meandering twisted pair made from 41.3 Ω/m Nichrome wire, was placed in the central layer of the stack (Figure 1, see (b)-(d)). Either 5 (Set 1) or 7 (Set 2) layers of YBCO were used on each side of the central heating strip. Thermocouples (E-type) were then distributed throughout the stack. There were two sets of measurements, Set 1 included all the ZnO-based stacks, Set 2 included the Nomex, Kapton, and Mylar based stacks. For Set 1, five

YBCO strips were used on each side of the heater, and thermocouples T3-T5 were used for the thermal measurements. For Set 2, seven strips of YBCO were used on each side of the heater. Thermocouples were labeled and measured on one side of the heater only; they were denoted T1-T5.

The sample segment was then placed into a block of closed cell styrofoam with a 4 cm by 0.4 cm opening designed to allow thermal diffusion in one dimension only (Figure 1(a)). The thermocouple readings were taken using a Keithley Integra series 2700 multimeter set up to read E-type thermocouples. The heating strip was connected in series with an HP 6114A precision power supply and an HP 3457A multimeter/ammeter. The sample edges were sealed to the styrofoam with silicone grease to prevent liquid nitrogen entry and to thus maintain the validity of the 1-D approximation.

## 2.2. Preparation of the Coil and Instrumentation for Coil Thermal Diffusion and Quench Measurement

Three YBCO pancake coils, designated A, B, and C, were wound with conductor 2, which was 4 mm wide and surrounded on all sides by a 22-μm thick copper stabilizer. The specified minimum $I_c$ was 60 A at 77 K and self-field. All used G10 coil formers 3.5 cm high, and with an outer former diameter (coil ID) of 7.62 cm. The YBCO coils were vacuum impregnated using CTD101 epoxy, this is summarized in Table 2. During the winding, heater segments, thermocouples, and voltage taps were introduced. For Coil A, YBCO tape was co-wound with a 19-μm thick Mylar tape. For Coils B and C, the YBCO tape was sputter-coated with 200 nm thick ZnO at the National Renewable Energy Laboratory (NREL). This coating served as the insulation, replacing the Mylar.

In preparation for measurement the room temperature resistance per unit length of each coil was measured to insure no layer-to-layer shorting was present. Additionally, $I_c$ (77 K) was measured for each segment of the coils to insure that the tape was not damaged in coil winding. After this, radial thermal conductivity and quench propagation were measured for each coil while immersed in $LN_2$. Radial thermal gradients were measured by applying a constant power to the heater (no applied current), and then measuring the temperature at various positions in the winding using the embedded E-type thermocouples. Quench propagation was measured by applying a set current (a DC current at some fraction of $I_c$), after which the heater was energized at a set current for a given time. Quench propagation was then monitored by the voltage developed between pairs of taps embedded in the winding. The system for quench propagation measurement consisted of an A/D board whose amplified signal was read by a computer using Labview. There were sixteen input channels, each with a 16 bit resolution on the A/D board, +/- 5V full scale. The input amplifiers had a gain that could be varied from 1 to 2000, giving an input resolution of 10V/2000/65536 = ~0.076 µV, or about 80 nV. The response time was limited by the 10 kHz filter at the input of the amplifier, leading to a time resolution of 0.5 milliseconds.

### 3.0 Thermal Propagation Measurements on the Short Stacks

Thermal gradients were measured on six different YBCO short sample stacks forming two sets (three samples each). The first set of samples (S1-S3) had ZnO based insulations, the second set (S4-S6) had Nomex, Kapton, and Mylar insulations. Figure 2 shows the thermal gradients in short sample stacks S1–S3. Per layer $\Delta T$ (defined as the temperature difference between a pair of thermocouples divided by the number of YBCO/insulation layers in-between them) is plotted vs heater power in watts. The measurements were performed on thermocouples T4-T5. The thermocouples are distributed as shown in the inset, which shows one half of the stack with the heater at the top of the diagram (above layer 1), and the liquid nitrogen bath at the bottom (below layer 5). Table 3 lists some of the physical parameters of the stacks.

The experimentally measured thermal conductivity for the YBCO stacks were extracted from the standard equation for heat diffusion, namely

$$\Delta T = \frac{QL}{A\kappa} \tag{1}$$

where $\Delta T$ is the temperature difference (in kelvins), $Q$ is the heat transported (in watts), $L$ is the distance (in meters), $A$ is the area (m$^2$) for heat flow, and $\kappa$ is the thermal conductivity (Wm$^{-1}$K$^{-1}$) of the composite stack. Since in our case, the stack is exposed to the nitrogen on both wide faces, the heat flow in Eq (1) goes both ways, and thus the total cooling area is twice the area of the stack face. The appropriate length is ½ of the total stack depth. The results (averaged over both T3-T5 and T4-T5) are presented in Table 3; note that $\kappa$ for the Cu doped ZnO is slightly higher than the others.

Figure 3, paralleling the data of Figure 2, shows the thermal gradients for short sample stacks S4–S6 (Nomex, Kapton, and Mylar). Per layer $\Delta T$ is again plotted vs heater power in watts. Measurements were performed on thermocouples T2-T5, with thermocouples placed as shown in the inset. The overall thermal conductivities are similar to those of S1-S3, which suggests that the high thermal conductivities possible for ZnO based materials are either not present, or another part of the composite is limiting $\kappa$. The values for $\kappa$ are average values across the whole stack, rather than specific to the insulation itself. However, it is possible to more closely estimate the actual layer thermal conductance by calculation, as shown below.

### 3.1. Estimation of the Thermal Conductance

The values measured above for $\kappa$ are composite values. We can use a simple set of calculations to estimate a composite (average) value based on the properties of each of the composite elements. This will allow us to; (i) predict the final composite value from that of the constituent elements, or (ii) extract the value of a given element if the average is measured and all other components are known or estimated. There are two main directions to be considered, along the conductor (or azimuthally within the coil), and perpendicular to the conductor (or radially within the coil).

***Tape Perpendicular (Radial) Thermal Conductance***

For the tape perpendicular (radial) thermal conductance the elements must be treated as a series conductance, as given by

$$\langle \kappa \rangle = \ell \left( \sum_i \frac{\ell_i}{\kappa_i} \right)^{-1} \tag{2}$$

Here $l$ is the total perpendicular distance averaged over, $l_i$ is the perpendicular length of a given element, and $\kappa_i$ is the thermal conductivity of the $i^{th}$ element. We assume here that the $i^{th}$ element has an isotropic conductivity, although in principle anisotropy could easily be accounted for.

First, we start with the tape itself, some relevant parameters of which are given in Table 4. Then putting in numbers (see Table 4, and ref [17-23]), and ignoring for now the edges of the tape, we obtain, for conductor type 2,

$$\langle \kappa \rangle_{\perp, tape} = 98 \mu m \left( \frac{50 \mu m}{7 W m^{-1} K^{-1}} + \frac{44 \mu m}{520 W m^{-1} K^{-1}} + \frac{2 \mu m}{430 W m^{-1} K^{-1}} + \frac{2 \mu m}{8 W m^{-1} K^{-1}} \right)^{-1} \tag{3}$$

where $\kappa_{\perp, tape}$ is the tape perpendicular conductivity ignoring the edge effects. Once evaluated, this gives 13.1 $Wm^{-1}K^{-1}$ for thermal conduction through the tape. Investigating the influence of each term separately, we will find that the dominant term (the first in Eq (3)) is the substrate. In fact, if we simply assume that the other layers have infinite conductivity, we would obtain 13.7 Wm⁻¹K⁻¹, suggesting that to a good approximation the thermal conductivity is just a kind of length averaged $\kappa$ of the substrate.

However, if we add in the parallel portion of Cu at the edges, we get (for Conductor 2)

$$\langle \kappa \rangle_{\perp, tape+e} = 13.1 \quad Wm^{-1}K^{-1}\left(\frac{4000}{4044}\right) + 520 Wm^{-1}K^{-1}\left(\frac{44}{4044}\right) = 18.6 \quad Wm^{-1}K^{-1} \tag{4}$$

where $\kappa_{\perp, tape+e}$ is the tape perpendicular conductivity including the edge effects. These values are listed in Table 4, along with similar calculations for Conductor type 1.

We can now calculate the case of a winding or stack layer, including the conductor layer, the interlayer (insulation + epoxy), and also the effects of the conductor edges. Approximating the conductor as rectangular, we can see the average layer (conductor layer + interlayer) conductivity as two parallel channels, one through the tape, and one through the tape edges. Each of these components is a series average of conductor layer and interlayer contributions, leading to

$$\langle \kappa \rangle_{\perp layer} = \frac{w_e \kappa_e + w_{cen} \kappa_{central}}{w_e + w_{cen}} = \frac{1}{w_e + w_{cen}}\left\{ w_e L_{layer}\left(\frac{d}{\kappa_{Cu}} + \frac{L_{int}}{\kappa_{int}}\right)^{-1} + w_{cen} L_{layer}\left(\frac{d}{\kappa_{\perp tape}} + \frac{L_{int}}{\kappa_{int}}\right)^{-1} \right\} \tag{5}$$

Here $\kappa_{\perp Layer}$ is the average layer thermal conductivity in the tape perpendicular direction, $w_e$ and $w_{cen}$ are the widths of the strand in the edge (Cu only) and central regions, $\kappa_e$ is associated with the edge regions, and $\kappa_{central}$ the central regions of the layer. $L_{Layer}$ is the total layer thickness = $d$ + $L_{int}$, and $\kappa_{int}$ is the average thermal conductivity of the epoxy and the insulation. We can then estimate what the $\kappa_{\perp layer}$ is for stacks or windings with various insulations. The values for stacks using Nomex, Kapton, and Mylar are listed in Table 4, assuming for simplicity negligible epoxy thickness, for both Conductor 1 and 2. The values for conductor 1 range from 0.46-1.57 Wm$^{-1}$K$^{-1}$. The measured results for the stacks are similar to this, if somewhat smaller ranging from 0.7-0.9 Wm$^{-1}$K$^{-1}$. We have also made an estimate for the stacks with ZnO-based insulations, this estimate is 14.9 Wm$^{-1}$K$^{-1}$for conductor type 1, essentially that of the conductor itself, because of the thinness and high thermal conductivity of the ZnO. The actual measured values are in fact quite different, ranging from 0.8 – 1.4 Wm$^{-1}$K$^{-1}$. The differences between experiment and measurement for the Nomex, Kapton, and Mylar insulations are within the realm of possible error contributions. However, the large deviation with the ZnO system is not, and suggests something else is limiting the thermal conductivity. Indeed, we must not forget the influence of the interface on the thermal conductivity, an effect referred to as thermal contact conductance.

A number of things contribute to thermal contact conductance, including both intrinsic physics effects, as well as the partial contact of microscopically rough surfaces, and the presence of native oxides or other contaminants. Taking only the influence of the native Cu oxides themselves, we see that the effects can be substantial. These layers, while even thinner than the ZnO layers (about 4-5 nm [24,25]) are quite resistive, both electrically [26-28] and thermally. Of course other factors also play a part making a de-convolution of the various effects difficult. Fortunately, some experimental results do exist. A number of studies have looked at heat flow between bulk metallic materials butted directly together, or with various interlayers, and with varying pressures, using thermal measurements performed at low temperatures [29], estimations can also be made from electrical measurements [30]. Based on these measurements, specifically for the case of epoxy interlayers, the thermal interface conductance can vary greatly, from roughly $R_H^{-1} = 10^3$ to $10^5$ Wm$^{-2}$K$^{-1}$, where $R_H$ is defined from [29]

$$\Delta T = Q\left(\frac{L}{A\kappa} + \frac{R_H}{A}\right) \qquad (6)$$

If we choose $10^4$ Wm$^{-2}$K$^{-1}$ as a representative value, and then use the above equations to estimate the thermal conductivity for a Cu interface with no insulation at all (including only the thermal boundary resistance) we can define an effective interface $\kappa$ by noting that $\kappa_{eff} = L_{eff}R_H^{-1}$, or just directly use Eq 6, to obtain an average thermal conductivity of about 1.5 W m$^{-1}$K$^{-1}$. This value is in fact quite similar to the experimental measurements of all samples in this work, both those of the film insulations, and those using ZnO insulations. These effects can be seen to dominate the actual thermal resistance of the ZnO itself, and would be comparable in size to our estimates of the contributions from the thermal resistances of the insulations, treated as bulk materials, as we have done in Table 4. Of course, it is possible to lump all of these influences together and treat the whole insulation layer as a thermal contact resistance as has been done for Kapton in Ref [31], the results of experimental measurements under various pressure conditions [31] are quite variable but overall consistent with our measured and estimated values. It turns out then, that good estimates of radial (tape perpendicular) thermal conductivities are difficult to make from first principles, suggesting that experimental measurements are essential.

***Tape Parallel (Azimuthal) Thermal Conductance***

For the tape parallel (azimuthal) thermal conductance the elements must be treated as a parallel conductance. Here we restrict ourselves to heat flow parallel to the tape, and in the direction of current flow within the tape, that is, along the azimuthal direction in a coil. The thermal conductance is then given by

$$\langle \kappa \rangle_{//,tape} = \left( \frac{1}{A_{cond}} \right) \left( \sum_i \kappa_i A_i \right) \tag{7}$$

Where $A_{cond}$ is the total area of the conductor end-on, $A_i$ is the end on area of the $i^{th}$ component of the conductor, and $\kappa_i$ is the $i^{th}$ thermal conductivity. Using the physical dimensions and thermal conductivities in Tables 1 and 4, we can find $\kappa$ to be 163 $Wm^{-1}K^{-1}$ and 261 $Wm^{-1}K^{-1}$ for Conductors 1 and 2, respectively. We can then again use a parallel treatment, in this case of the conductor and the insulation, to obtain the average tape parallel (azimuthal) conductivity, which is given by

$$\left\langle \kappa \right\rangle_{//,layer} = \left( \frac{1}{A_{layer}} \right) \left( A_{cond} \kappa_{cond} + A_{ins} \kappa_{ins} \right) \tag{8}$$

The values $\kappa_{//,layer}$ with Nomex, Kapton, Mylar, and ZnO based insulations are listed in Table 4. As a practical matter, the thermal conductivity contribution of the insulation layer is negligible, and the final layer thermal conductivity is an area normalized $\kappa_{//,cond}$.

### *Coil Axial Thermal Conductance*

Now, we must also consider the thermal conductance axially within the coil, which corresponds to heat flow parallel to the wide face of the tape, but perpendicular to the tape winding direction (or current flow direction). The thermal conductance is again a parallel conductance calculation, and gives in fact the same value as the above tape parallel, azimuthal conductance, to a very good approximation. This does assume that the coil is wound as a single pancake coil, or a series of pancakes coil, with negligible thermal barrier between them. If we wished to treat stacked pancake coils with non-negligible thermal barriers, we would need to modify the calculations accordingly, but this is not treated in this work.

## 4.0 Coil Measurements

### 4.1. Measurements of Coil A (1 m/4 turns, Mylar Insulated)

For Coil A, Mylar insulation was used, and it was co-wound with the YBCO tape. The total tape length was 1 m, leading to a total of four turns. Two heaters were placed in this coil (HT1 and HT2), as shown in Figure 4. The primary heater, HT1, was 12.7 cm long with a resistance of 300 Ω, while a backup heater, HT2, was 6.35 cm long, with $R$ = 260 Ω. Three thermocouples (T1-T3) were used; the radial distance T1-T2 and T2-T3 was 0.22 mm in each case, such that between T1 and T3 the total distance was 0.44 mm. Five voltage taps were also embedded in the coil, with a fixed distance of 14.0 cm between each tap and its neighbor, except for the distance between V5 and V6, which was 12.7 cm. Resistance measurements were made between all sequential pairs of voltage taps at room temperature to check for shorts, none were found. Transport $I_c$ measurements were also made between all pairs of sequential voltage taps (injecting current at the current leads), and the $I_c$ values ranged from 49.2 to 69 A.

### 4.1.1 Thermal Gradient Measurements of Coil A (1 m/4 turns, Mylar Insulated)

Figure 5 (a) shows the temperature vs heater power in W/cm$^2$ at thermocouple positions T1-T3 along the radial direction of Coil A at heater powers of 0.752 W (0.148 W/cm$^2$), 1.69 W (0.322 W/cm$^2$), and 3 W (0.591 W/cm$^2$). No transport current is flowing for the thermal gradient measurements. Here, $T_p$ = $T_h$ - $T_0$ (the temperature at a given heater power, $T_h$, minus the temperature with no applied power, $T_0$). Figure 5 (b) shows the temperature gradient radially across Coil A for these heater powers. These data can be used to evaluate the radial thermal conductivity of the winding. However, because these data are steady-state, they reflect the heat conduction both along the tape tangentially as well radially, so a model which accounts for this must be used.

Here we use a 1D model for the steady state heat conduction of a hollow cylinder, namely

$$\kappa = \frac{Q \ln\left(\dfrac{R_2}{R_1}\right)}{2\pi L (T_1 - T_2)} \tag{9}$$

where $\kappa$ is the thermal conductivity of the winding (in $Wm^{-1}K^{-1}$), $Q$ is the rate of heat flow (in Watts), $R_1$ and $R_2$ are the inner and outer radius of the winding, respectively, $L$ is the axial length of the winding (in this case the width of the tape), and $T_1$ and $T_2$ are the temperatures on the inner and outer surface of the winding, respectively. Under this 1-D assumption (that all the heat generated by the heater propagates across the winding) an average thermal conductivity can be obtained for the winding, for Coil A this is $\kappa = 1.55$ $Wm^{-1}K^{-1}$. This value is relatively close to the estimation for $\kappa_{Llayer}$ above, at 1.22 $Wm^{-1}K^{-1}$, and somewhat higher than the measured values for the stacks, at 0.9 $Wm^{-1}K^{-1}$. However, overall the agreement between the two measurements and the expectation is reasonable.

.

### 4.1.2. Quench Propagation Measurements for Coil A (1 m/4 turns, Mylar Insulated)

Figure 6 shows the quench response for Coil A. First, a steady transport current of 34.8 A was applied to the coil, this value was 0.7 of the minimum $I_c$ of the coil which was 49.2 A. To initiate the quench, heat pulses of 0.59 $W/cm^2$ (heat current = 100 mA) were then applied for various times using HT1. Under these conditions, and with heater pulses lasting 5 – 7 s, only region V1-V2 developed a normal zone. For pulse durations of 8 s or more, a normal zone appeared in region V3-V4, and at 9 s it appeared also between V5-V6. Figure 6 shows this last condition, with a 9 second heat pulse applied. For 10 s pulse duration, V6-V7 also developed a normal zone. The normal regions appeared in the following order: (1) V1-V2, (2) V3-V4, (3) V5-V6, and for 10 sec pulses (V6-V7). The fact that the voltage taps do not quench in a consecutive order excludes the possibility of normal zone propagation along the tape length and suggests its propagation in a radial direction across the winding. This is caused by two facts: (i) there is a significantly smaller cross-section for heat transfer in azimuthal direction, i.e. longitudinally along the tape (given by the cross-section of the tape $w$ x $t$) compared with the cross-section in the radial direction ($w$ x 1 turn length), even though the thermal conductivity along the tape is higher than across the winding, (ii) the distance over which the normal zone needs to propagate radially in order to get to the next layer is quite small because of the thinness of the tape, while a azimuthally (longitudinally) propagating normal zone must travel around the coil's circumference.

The coil was not damaged during this series of quench experiments; we repeated the measurements with a heat pulse of 99.7 mA for 10 s and we obtained results nearly identical with those from the previous runs.

## 4.2. Measurements of Coil B (1 m/4 turns, ZnO Insulated)

Coil B used a ZnO coating applied directly to the tape as the insulation. The total tape length was 1 m, and there were four turns in the coil. Two heaters were placed in this coil (HT1 and HT2), as shown in Figure 7. The main heater, HT1, was 12.7 cm long and had a resistance of 640 $\Omega$, while the backup heater, HT2, was 6.35 cm long with $R = 581 \Omega$. Three thermocouples (T1-T3) were used. Seven voltage taps were distributed uniformly in angle (to improve our ability to assess the now-expected radial propagation) as shown in Figure 7. Along the radial direction, the voltage taps were 1 layer apart, and within the coil they were 120° apart. Resistance measurements were made between all sequential pairs of voltage taps at room temperature to check for shorts, none were found. Transport $I_c$ measurements were also made between all pairs of sequential voltage taps (injecting current at the current leads), and the $I_c$ was relatively homogeneous with an average $I_c = 73.8 \pm 5.8\%$.

### 4.2.1. Thermal Gradient Measurements of Coil B (1 m/4 turns, ZnO Insulated)

Figure 8 (a) shows the temperature profile for Coil B, in terms of $T_p$ vs heater power for several different power inputs. Here, $T_p = T_h - T_0$ (the temperature at a given heater power, $T_h$, minus the temperature with no applied power, $T_0$). Figure 8(b) shows the temperature gradient at two different power levels. We again can use Eq (9) to extract $\kappa$; obtaining $\kappa = 2.79$ Wm$^{-1}$K$^{-1}$. In this case, only the outer two thermocouple layers were considered. We do notice in Figure 8 that there is a substantial difference in the gradients between T1–T2, as compared to T2-T3. This is most likely due to differences in the axial position (height) of the thermocouples. This is because even though we have considered the coil to be a cylinder uniform in temperature azimuthally, consistent with its high azimuthal thermal conductivity, we must also consider its high axial thermal conductivity. While this thermal conductivity is much higher than that perpendicular to the tape, if the thermocouples are all at the same vertical position, the measurement of $\kappa_\perp$ will be unaffected. However, axial displacements will introduce errors, since there will be an axial temperature gradient. This effect is reduced by the influence of the epoxy coating on the edges of the pancake coil, but some errors remain for the data in Figure 8, suggesting some displacement of the thermocouples axially. In any case the value of $\kappa = 2.79$ $Wm^{-1}K^{-1}$ is consistent with measurements made on coil C, below, and can be compared to the experimental measurements of 0.8-1.4 $Wm^{-1}K^{-1}$, and the theoretical estimation of 13.1 $Wm^{-1}K^{-1}$. As the results for the coil measurement are again much lower than the "expected" values, we again see the influence of an additional thermal barrier, as described above.

### 4.2.2. Quench Propagation Measurements for Coil B (1 m/4 turns, ZnO Insulated)

Figures 9-11 show the quench propagation through Coil B (1 m, ZnO insulation) with increasing energy deposition. First, Figure 9 shows a quench propagation measurement where $I = 34.8$ A (1/2 of the minimum $I_c$ of 69.5 A) was supplied to the coil continuously, then at about $t = 2$ seconds, HT1 was excited with 75 mA, giving 680 mW/cm$^2$, for 5 seconds. Normal zones are seen to form between V1-V2, V4-V5 and V6-V7, regions of the coil at the same angular position, but at increasing radial distances. The measurements thus indicate radial heat propagation. A second set of measurements are shown in Figure 10, in this case at a higher heat pulse level. In this run $I = 34.8$ A was supplied to the coil continuously, then at $t = 0$, HT1 was excited with 100 mA, giving 1.21 W/cm$^2$, for 2 s. Normal zones are seen to form for V1-V2, V4-V5, V6-V7 and V3-V4. Taps V1-V2, V4-V5 and V6-V7 are positioned at the same azimuthal position as HT1 at increasing radial distances within the winding. The measurements again indicate radial heat propagation. A third set of quench measurements for Coil B are shown in Figure 11, in this case at a higher current level. In this run $I = 54.1$ A (0.78 of the minimum $I_c$ of 69.5 A) was applied to the coil continuously, then at $t = 0$ HT1 was excited with 100 mA, giving 1.21 W/cm$^2$, for 160 s (12 seconds of which are shown). Normal zones are seen to form in the order V1-V2, V4-V5, V3-V4, V6-V7, V5-V6, and finally V2-V3. A partial recovery was seen in taps V3-V4 and V5-V6 after 1.5 seconds.

Figure 12 shows the NZP radial velocities ($NZP_{ra}$) for a series of quench measurements performed on Coil B. There are three grouping, the first set of runs was performed with a DC coil current of 34.7 A and a heater deposition of 3.47 W (680 mW/cm$^2$), the second set at $I =$ 34.7 A and a heater power of 6.16 W (1.21 W/cm$^2$). Within these sets the results for different heater excitation times are shown. The last two columns show data for elevated DC currents in the sample (42.37 A (61% of $I_{c,min}$) and 54.12 A (78% if $I_{c,min}$)), applying 6.16 W for 160 seconds. Values of $NZP_{ra}$ were typically in the range 0.2~0.4 mm/s. There was no systematic dependence of $NZP_{ra}$ on heat pulse duration. However, there was the expected dependence on heat pulse power as well as on the transport current of the sample.

Finally, not shown in the present set, further measurements were made which put the azimuthal NZP propagation at about 17 mm/s.

## 4.3. Measurements of Coil C (12 m/45 turns, ZnO Insulated)

Coil C also used a ZnO insulation. The total tape length was 12 m, leading to a total turn number of 45, see Figure 13. The primary heater, HT1 had a resistance of 666 $\Omega$, and a length of 12.7 cm. Five thermocouples (T1-T5) were placed in the angular center of HT1 with 1.53 mm between each of them in the radial direction . Thirty voltage taps were distributed uniformly in angle as shown in Figure 13. Along the radial direction, the voltage taps were 10 layers apart, and along the azimuthal direction they were 60° apart.

### 4.3.1. Thermal Gradient Measurements of Coil C (12 m/45 turns, ZnO Insulated)

Thermal gradients for Coil C are shown in Figure 14 (again, $T_p = T_h - T_0$). Using again Eq (9) and the data in Figure 14 we obtained an average thermal conductivity of the winding of Coil C as $<\kappa>=2.35$ Wm$^{-1}$K$^{-1}$. Here we took the gradient from the steepest portion of the curve, that between T1-T2, assuming at this low level of excitation the heat is removed by the T3 layer. This value can be compared to the previous coil measurement, Coil B, which yielded $\kappa = 2.79$ Wm$^{-1}$K$^{-1}$ as well as to the previously mentioned experimental results on stacks (0.8-1.4 Wm$^{-1}$K$^{-1}$, Table 3) and the theoretical estimate of 13.1 Wm$^{-1}$K$^{-1}$ (Table 4).

### 4.3.2. Quench Propagation Measurements for Coil C (12 m/45 turns, ZnO Insulated)

Figures 15 and 16 shows the quench properties of Coil C (45 turn/12 m, ZnO insulation). Due to the limited number of channels on our data acquisition device, we chose to look at the "upper" portion of the coil as shown in Fig 13 (top); that is voltage taps on the upper portion of this figure were attached to the data acquisition device. Figure 15 shows a quench measurement where $I = 34.9$ A (50% $I_{c,min}$) was applied to the coil continuously, then at $t = 0$, HT1 was excited with 100 mA, giving 1.21 W/cm$^2$, for 60 s. The data has been smoothed with a lowpass filter for clarity. A normal zone is shown for V1-V6, normal zones also form (in sequence) for V13-V18 and V10-V11, V19-V24, and V25-V30. All of these except V10-V11 are at increasing radial distances above each other and the heater, indicating again radial NZP. This NZP$_{ra}$ = 0.33 mm/s. The generation of a normal zone on taps V10-V11 indicates an azimuthal normal zone propagation, the onset occurs at 3 s, yielding NZP$_{az}$ = 13.8 mm/s, comparable to the 17 mm/s measured for Coil B.

Figure 16 shows a quench measurement where $I = 44.53$ A (64% $I_{c,min}$) was applied to the coil continuously, then at $t = 0$ seconds, HT1 was excited with 100 mA, giving 1.30 W/cm$^2$, for 5 s. We then follow the progress of the normal zone formation with no heater excitation from $t = 5$ seconds to 50 seconds. A normal zone is shown for V1-V6, and then normal zones also form for V13-V18, V19-V24, and V25-V30. Here again, the propagation is predominantly radial. The instrumentation level of this coil allows us to obtain good numbers for the radial and azimuthal NZP, viz. 0.33 m/s and 13.8 mm/s, respectively.

## 5.0 Discussion

In this work, three different insulations types were investigated and compared to more conventional Nomex, Kapton, and Mylar insulations. The thermal conductivities of the composite stacks made with these insulations (under 8.9 MPa and with stycast epoxy) were measured and found to be in the range of $0.8 - 1.4$ Wm$^{-1}$K$^{-1}$ – not very different from similar stacks made using the more conventional insulations. In fact the estimated interface thermal contact resistance values are of a similar level to the actual measurements, and likely dominate the contributions from the ZnO layers in the associated stacks. These layers probably also contribute substantially to the effective thermal conductance even for the more conventional insulation packs. On the other hand, the actual size of the layers was reduced significantly with the use of ZnO insulation (about 20%). In addition, the thermal propagation level in the coils was seen to be somewhat better for the ZnO insulations, improving by some 50-100%. The difference between the single stack results and the coil results may have had to do with the different levels of layer compaction in the two, even though an effort was made to make them similar.

Three pancake coils were then fully measured, two of which were 1 m long and had Mylar and ZnO insulations, respectively, and one final coil with ZnO insulation which was 12 m long. The coils were measured for their thermal conductivity and normal zone propagation velocity. The results for $\kappa$ via coil measurement, stack measurement, and calculation, are shown in Table 5. Coil measurements give a $\kappa$ value about twice that of the stack measurements, for reasons that are not clear. The calculated value for $\kappa$ is roughly similar for he mylar insulated coil, but do not agree for the ZnO insulated coils because we have neglected the interface contribution (which is difficult to estimate). It can be seen that the thermal conductivities are slightly higher for the ZnO as compared to the Mylar (50-100%).

The NZP values both radially and azimuthally are shown in Table 5. The azimuthal values, 14-17 mm/s, are similar to those seen for single coated conductor tapes [1-5]. The radial values, of course, are much smaller, 0.05-0.3 mm/s, reflecting the much lower $\kappa$ value in that direction. The most interesting point however, is that the propagation of the NZP is in fact radial. This is only because of the thinness of the YBCO tape as compared to the perimeter of a winding, the ratio of winding length to thickness is 16 cm/0.01 cm = 1600 (as compared to the longitudinal and radial NZP values which have a ratio of ~ 100. Thus, due to geometry effects (thinness of YBCO film as compared to much larger perimeter of the coil), one should expect the layer-by-layer propagation to be faster than the longitudinal propagation.

**Summary**

The thermal diffusion properties of several different kinds of YBCO insulations and the quench properties of coils made using these insulations were studied, specifically comparing a high thermal conductivity insulation (ZnO) to Nomex, Kapton, and Mylar insulations. One of the goals was to increase the radial speed of propagation for the normal zone, in order to compensate for the relatively low longitudinal NZP for YBCO, thus making the coils easier to quench protect. However, the results for all short sample and coil thermal conductivities were ~1-3 $Wm^{-1}K^{-1}$. The lack of distinction for the ZnO-based insulations was attributed to the presence of thermal interface contact resistance. On the other hand, the ZnO insulations, while not strongly increasing the average thermal conductivity of the winding pack or coil, were much thinner than the other insulations, and would thus enable substantial increases in winding pack critical current density.

Additionally, quench propagation velocity measurements were performed on three coils (77 K, self field). Normal zone propagation velocity (NZP) values were obtained for the coils both in the radial direction and in the azimuthal direction. Radial NZP values (0.05-0.7 mm/s) were two orders of magnitude lower than axial values (~14-17 mm/s). Nevertheless, the quenches were generally seen to propagate radially within the coils, in the sense that any given layer in the coil is driven normal by the layer underneath it. This initially surprising result is due to the fact that while the radial normal zone propagation velocity (NZP) is much lower than the NZP along the conductor (~100 x) the distance the normal zone must expand longitudinally is much larger than what it must expand radially to reach the same point, in our case this ratio is ~ 1600.

**Acknowledgements**

We thank David Ginley (NREL) for applying the ZnO based insulations to the YBCO tapes. This material is based upon work supported by the AFOSR under Contract No. FA9550-

## List of Tables

Table 1. Conductor Specifications.

|  |  | Conductor 1 | Conductor 2 |
|---|---|---|---|
| $I_c, A$ | Minimum $I_c$ | 95 | 60 |
| $w, mm$ | Conductor Width | 4.04 | 4.044 |
| $w_e, mm$ | Total conductor edge[a] width ($2d_{cu}$) | 0.04 | 0.044 |
| $w_{cen}, mm$ | Conductor central region width | 4 | 4 |
| $d_{ybco}, \mu m$ | YBCO thickness | 3 | 2 |
| $d_{Cu}, \mu m$ | Copper stabilizer layer thickness[a] | 40 | 44 |
| $d_{Ag}, \mu m$ | Silver overlayer thickness | 2 | 2 |
| $d_{Sub}, \mu m$ | Substrate thickness | 100 | 50 |
| $d, \mu m$ | Total conductor thickness | 145 | 98 |

[a] The tapes are fully enclosed (surrounded) by a $d_{cu}$ thickness of Cu.

Table 2. Coil Description

| Coil Name | Tape length, m | Conductor type | No. turns | Insulation | Coil ID, cm | Active coil height, mm | Former height, cm |
|---|---|---|---|---|---|---|---|
| A | 1 | 2 | 4 | Mylar | 7.62 | 4 | 3.5 |
| B | 1 | 2 | 4 | ZnO | 7.62 | 4 | 3.5 |
| C | 12 | 2 | 12 | ZnO | 7.62 | 4 | 3.5 |

Table 3. Measured Short Stack Thermal Conductivities

| Stack Name | Insulation | Insulation thickness, $t_{ins}$, μm | Epoxy | Tape | Conductor thickness, $t_{cond}$, μm | Pack half thickness, mm | No. YBCO layers | Av. Thermal conductivity, $<\kappa>$, Wm$^{-1}$K$^{-1}$ |
|---|---|---|---|---|---|---|---|---|
| *Set 1* | | | | | | | | |
| S1 | $Zn_2GeO_4$ | 0.69 | Stycast[a] | 2 | 98 | 0.62 | 10 | 0.8 ± 0.1 |
| S2 | ZnO-Cu | 0.69 | Stycast | 2 | 98 | 0.62 | 10 | 1.4 ± 0.3 |
| S3 | ZnO | 0.69 | Stycast | 2 | 98 | 0.62 | 10 | 0.8 |
| | | | | | | | | |
| *Set 2* | | | | | | | | |
| S4 | Nomex | 38 | Stycast | 1 | 145 | 1.33 | 14 | 0.7 ± 0.1 |
| S5 | Kapton | 40 | Stycast | 1 | 145 | 1.38 | 14 | 0.7 ± 0.1 |
| S6 | Mylar | 19 | Stycast | 1 | 145 | 1.60 | 14 | 0.9 ± 0.2 |

[a] Stycast 1266

Table 4. Thermal Conductivities and Dimensions.

| Component | Conductor 1 | | Conductor 2 | |
|---|---|---|---|---|
| | Thickness (μm) | $\langle\kappa\rangle$ (Wm$^{-1}$K$^{-1}$) at 77K | Thickness (μm) | $\langle\kappa\rangle$ (Wm$^{-1}$K$^{-1}$) at 77K |
| Copper stab. | 40 | 520 | 44 | 520 |
| Silver over. | 2 | 430 | 2 | 430 |
| Hast sub. | 100 | 7.00 | 50 | 7.00 |
| YBCO | 3 | 8 | 2 | 8 |
| total | 145 | -- | 98 | -- |

| Insulation Material | Thickness (μm) | $\langle\kappa\rangle$ (Wm$^{-1}$K$^{-1}$) at 77K |
|---|---|---|
| Nomex | 38 | 0.1 |
| Kapton | 40 | 0.4 |
| Mylar | 19 | 0.16 |
| ZnO* | $\cong$0.2 | $\cong$62 |
| Stycast 1266 | | $\cong$0.2-0.4[b] |
| CTD 101 | | 0.37 |

| $\langle\kappa\rangle$ (Wm$^{-1}$K$^{-1}$) | $\perp$ tape | $\parallel$ tape | $\perp$ tape | $\parallel$ tape |
|---|---|---|---|---|
| Tape, no edge | 9.84 | -- | 13.1 | -- |
| Tape, edge | 14.9 | 163 | 18.6 | 261 |
| Stack, Nomex[a] | 0.46 | 129 | 0.35 | 188 |
| Stack, Kapton[a] | 1.57 | 128 | 1.29 | 185 |
| Stack, Mylar[a] | 1.22 | 144 | 0.93 | 219 |
| Stack ZnO | 14.9 | 163 | 18.6 | 261 |

[a]assuming conductor 1 and zero thickness epoxy.
[b] Value is an estimate based on Ref [18],[19].

Table 5. Summary Results.

| Coil/Stack | $\kappa_{\perp, meas, coil}$ Wm$^{-1}$K$^{-1}$ | $\kappa_{\perp, meas stack}$ Wm$^{-1}$K$^{-1}$ | $\kappa_{\perp theor}$, Wm$^{-1}$K$^{-1}$ | Radial NZP, mm/s | Azimuthal NZP, mm/s |
|---|---|---|---|---|---|
| Coil A | 1.55 | 0.9 | 1.22 | 0.05-0.1 | -- |
| Coil B | 2.79 | 0.8-1.4 | 13.1 | 0.1-0.7 | 17 |
| Coil C | 2.35 | 0.8-1.4 | 13.1 | 0.33 | 13.8 |

# Figure Captions

Figure 1. Short stack thermal measurement arrangement, including; (a) Short sample block mounted in thermal gradient measurement holder, (b) Short sample block, consisting of an epoxied stack of instrumented and insulated YBCO conductor segments, (c) Schematic position of thermocouples on different layers, and (d) in-plane arrangement of nichrome heater wire meander.

Figure 2. Thermal gradients in short sample stacks S1–S3. Per layer $\Delta T$ (defined as the temperature difference between a pair of thermocouples divided by the number of YBCO/insulation layers in-between them) is plotted vs heater power in Watts. Measurements performed on thermocouples T3-T5, with thermocouples placed as shown in inset. Inset shows one half of the stack, with the heater at the top of the diagram (above layer 1), and the liquid nitrogen bath at the bottom (below layer 5).

Figure 3. Thermal gradients in short sample stacks S4–S6. Per layer $\Delta T$ (defined as the temperature difference between a pair of thermocouples divided by the number of YBCO/insulation layers in-between them) is plotted vs heater power in Watts. Measurements performed on thermocouples T2-T5, with thermocouples placed as shown in inset. Inset shows one half of the stack, with the heater at the top of the diagram (above layer 1), and the liquid nitrogen bath at the bottom (below layer 7).

Figure 4. (a) Schematic of Coil A (1 m, Mylar insulated). Thermocouples are numbered T1-3, and voltage taps V1-V6. Two heater strips are located on the inner surface of the coil, HT1, the primary heater, and HT2, the backup heater, (b) Coil A after winding and epoxy impregnation.

Figure 5. Thermal gradients in Coil A (1 m mylar insulated); (a) $T_p$ (defined as the temperature at a given heater power, $T_h$, minus the temperature with no applied power, $T_0$) at T1, T2, and T3 vs heater power (HT1, $R = 300\ \Omega$, $L = 12.7$ cm) in W/cm$^2$, (b) Temperature gradient radially for various heater powers.

Figure 6. Quench propagation measurements for Coil A (1 m mylar insulated). In this run, $I = 34.8$ A was supplied to the coil continuously, then at $t = 0$, HT1 was excited with 100 mA, giving 590 mW/cm$^2$, for 9 s). Normal zones are seen to form between V1-V2, V3-V4, and V5-V6.

Figure 7. Schematic of Coil B (1 m, ZnO insulation). Thermocouples are numbered T1-3, and voltage taps V1-V7. Two heater strips are located on the inner surface of the coil, HT1, the primary heater, and HT2, the backup heater.

Figure 8. Thermal gradients in Coil B (1 m, ZnO insulation); (a) $T_p$ (defined as the temperature at a given heater power, $T_h$, minus the temperature with no applied power, $T_0$) at T1, T2, and T3 vs heater power (HT1, R = 616 $\Omega$, $L = 12.7$ cm) in W/cm$^2$, (b) Temperature gradient radially across Coil B for heater powers of 1.21 W/cm$^2$ (solid) and 3.17 W/cm$^2$ (dashed).

Figure 9. Quench propagation measurement for Coil B (1 m, ZnO insulation). In this run, $I$ = 34.8 A was supplied to the coil continuously, then at about $t$ = 2 seconds, HT1 was excited with 75 mA, giving 680 mW/cm$^2$, for 5 seconds. Normal zones are seen to form between V1-V2, V4-V5 and V6-V7, regions of the coil at the same angular position, but at increasing radial distances. The measurements thus indicate radial heat propagation.

Figure 10. A second set of quench propagation measurements for Coil B, in this case at a higher heat pulse level. In this run $I$ = 34.8 A was supplied to the coil continuously, then at $t$ = 0, HT1 was excited with 100 mA, giving 1.21 W/cm$^2$, for 2 s. Normal zones are seen to form for V1-V2, V4-V5, V6-V7 and V3-V4. Taps V1-V2, V4-V5 and V6-V7 are positioned at the same azimuthal position as HT1 at increasing radial distances within the winding. The measurements again indicate radial heat propagation.

Figure 11. A third set of quench propagation measurements for Coil B, in this case at a higher current level. In this run $I$ = 54.1 A was applied to the coil continuously, then at $t$ = 0 HT1 was excited with 100 mA, giving 1.21 W/cm$^2$, for 160 s (12 seconds of which are shown). Normal zones are seen to form in the order V1-V2, V4-V5, V3-V4, V6-V7, V5-V6, and finally V2-V3. A partial recovery was seen in taps V3-V4 and V5-V6 after 1.5 seconds.

Figure 12. NZP$_{ra}$ velocities for a series of quench measurements performed on Coil B. There are three grouping, the first set of runs was performed with a DC coil current of 34.7 A and a heater deposition of 3.47 W (680 mW/cm$^2$), the second set at $I$ = 34.7 A and a heater power of 6.16 W (1.21 W/cm$^2$). Within these sets the results for different heater excitation times are shown. The last two columns show data for elevated DC currents in the sample (42.37 A and 54.12 A),
applying 6.16 W for 160 seconds.

Figure 13. (a) Schematic of Coil C (45 turn/12 m, ZnO insulation). Thermocouples are numbered
T1-6, and voltage taps V1-V30. Two heater strips are located on the inner surface of the coil,
HT1, the primary heater, and HT2, the backup heater. (b) Coil C after winding and epoxy
impregnation.

Figure 14. Thermal gradients in Coil C (45 turn/12 m, ZnO insulation). $T_p$ (defined as the
temperature at a given heater power, $T_h$, minus the temperature with no applied power, $T_0$) vs
distance through the coil winding for 0.471 W/cm$^2$ (solid) and 1.31 W/cm$^2$ (dotted).

Figure 15. Quench propagation measurements for Coil C (45 turn/12 m, ZnO insulation). In this
run, $I = 34.9$ A was applied to the coil continuously, then at $t = 0$, HT1 was excited with 100 mA,
giving 1.21 W/cm$^2$, for 60 s. The data has been smoothed with a lowpass filter for clarity. A
normal zone is shown for V1-V6, normal zones also form for V13-V18 and V10-V11. NZP$_{ra}$ =
0.33 mm/s. On taps V10-V11 the onset occurs at 3 s, indicating NZP$_{az}$ = 13.8 mm/s. After this
normal zones on taps V19-V24, V25-V30 and V22-V23 start to become visible.

Figure 16. Quench propagation measurements for Coil C (45 turn/12 m, ZnO insulation). In this
run, $I = 44.53$ A was applied to the coil continuously, then at $t = 0$ seconds, HT1 was excited
with 100 mA, giving 1.30 W/cm$^2$, for 5 s. We then follow the progress of the normal zone
formation with no heater excitation from $t = 5$ seconds to 50 seconds. A normal zone is shown for V1-V6, normal zones also form for V13-V18, V19-24, and V25-30.  We note that the voltage formation in V25-V30 is more rapid than expected; the reason for this is unknown.

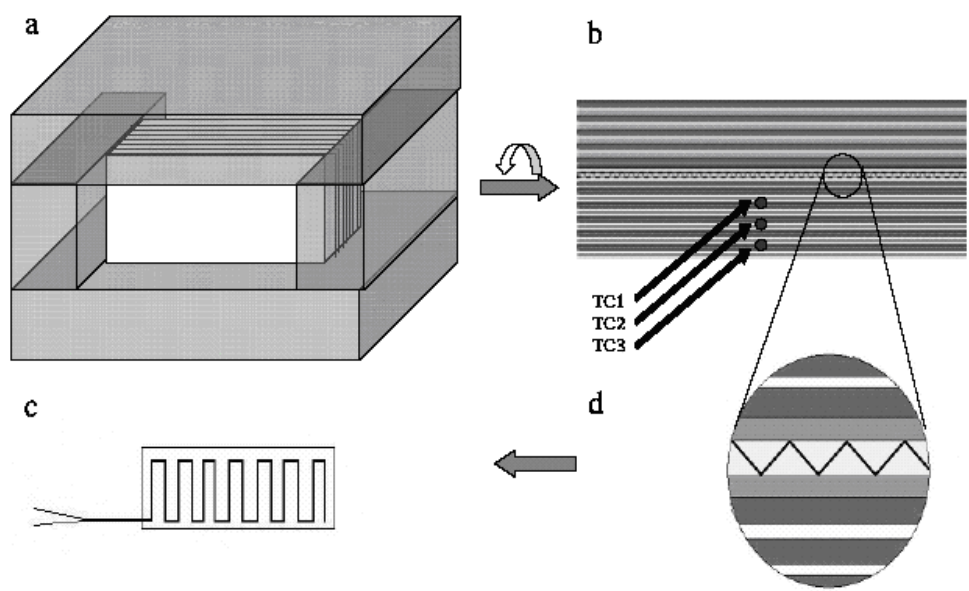

Fig. 1.

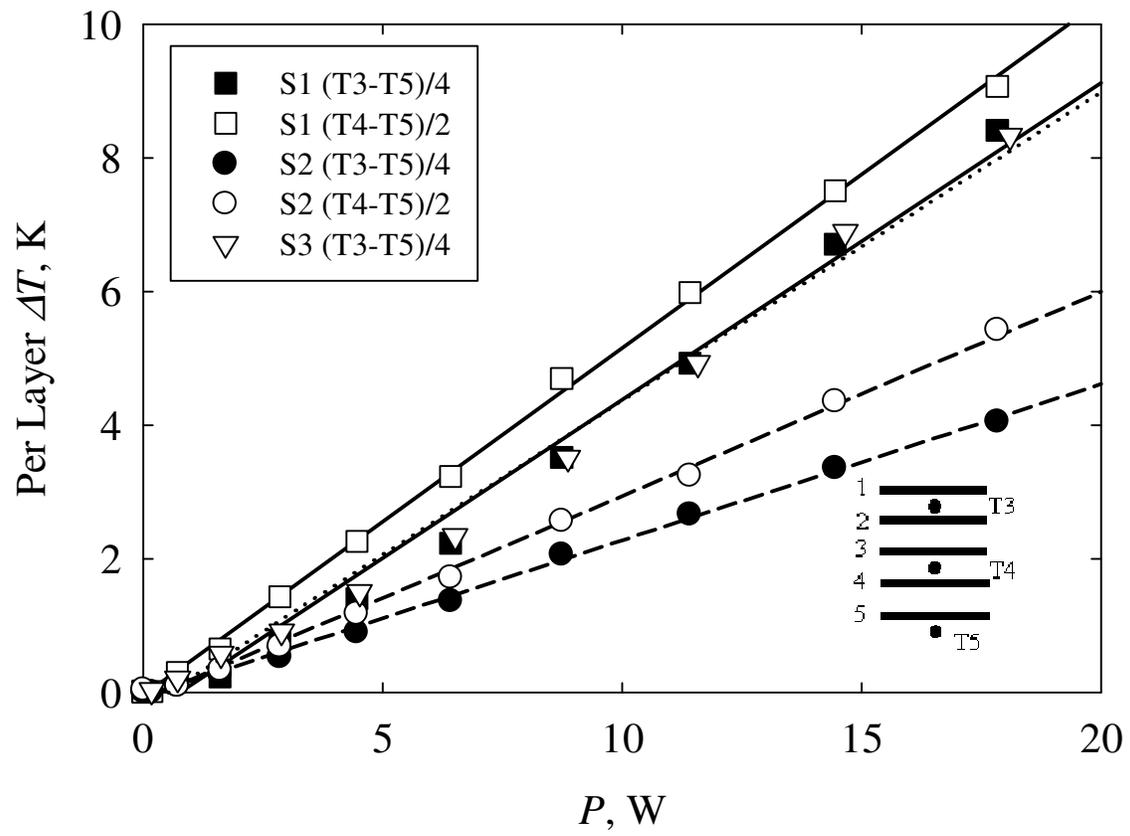

Fig. 2.

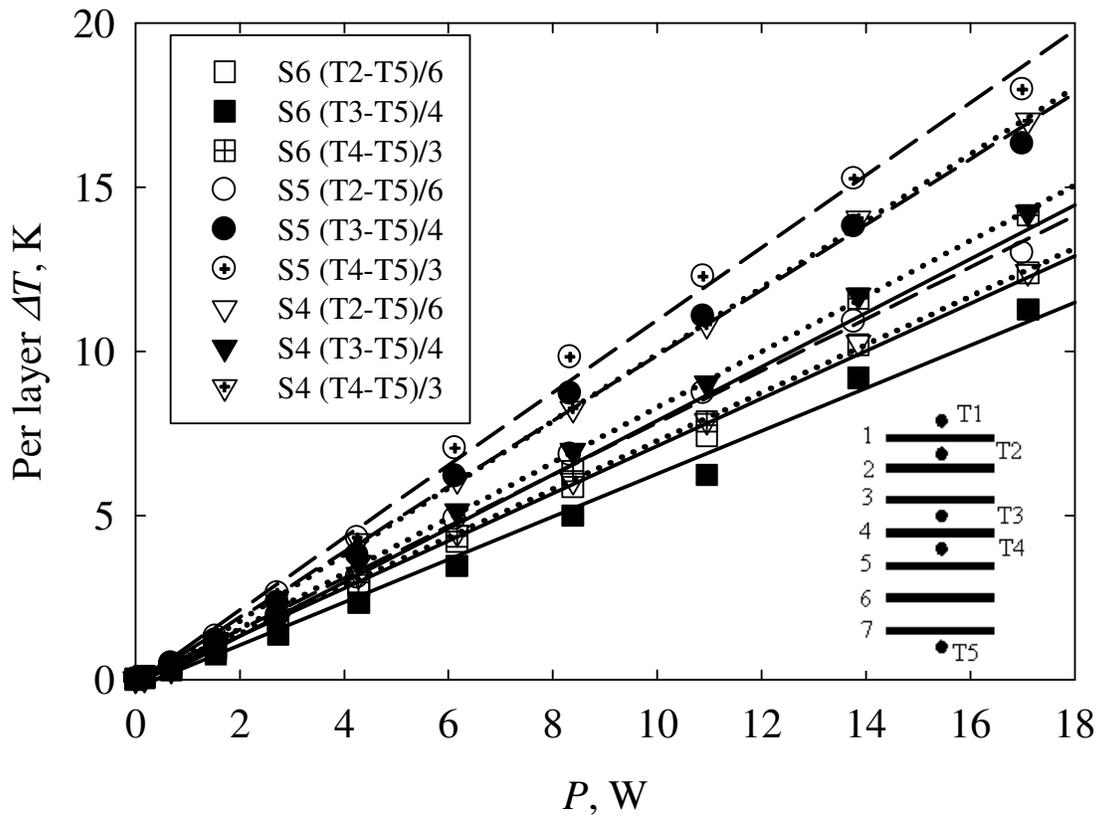

Fig. 3.

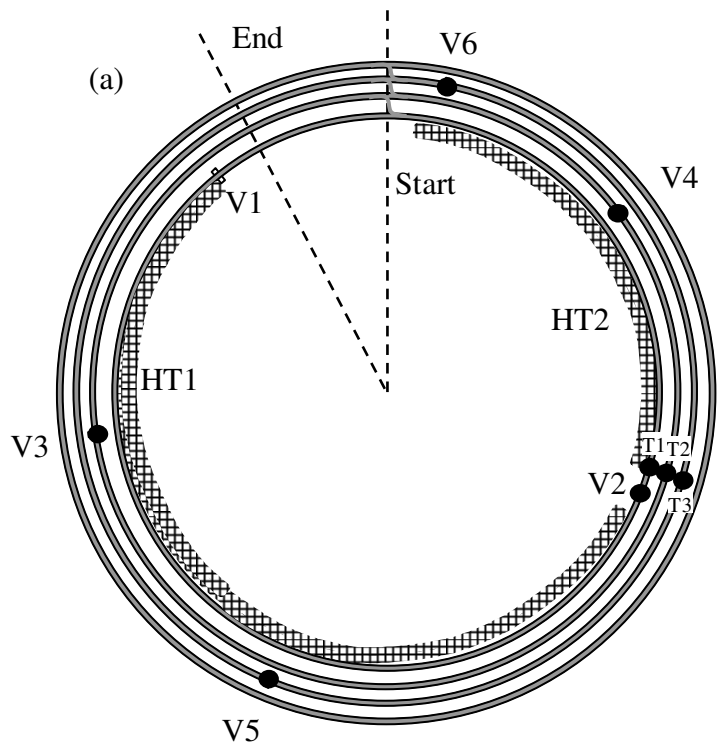

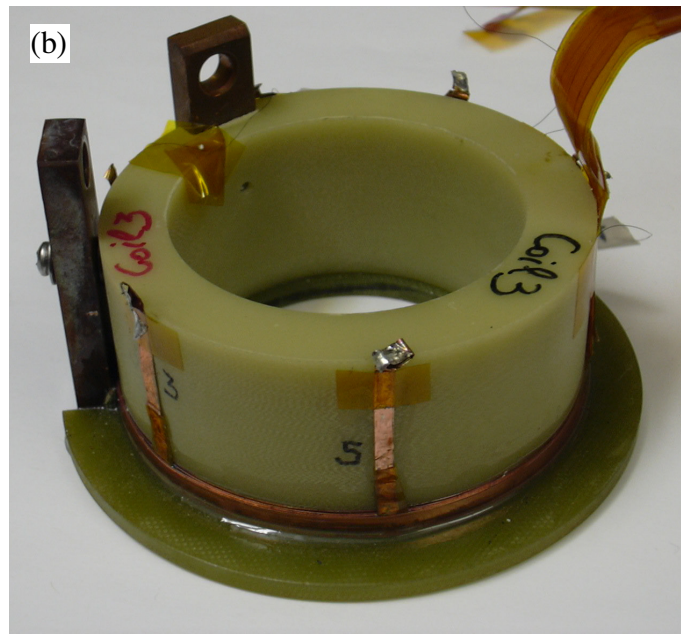

Fig. 4.

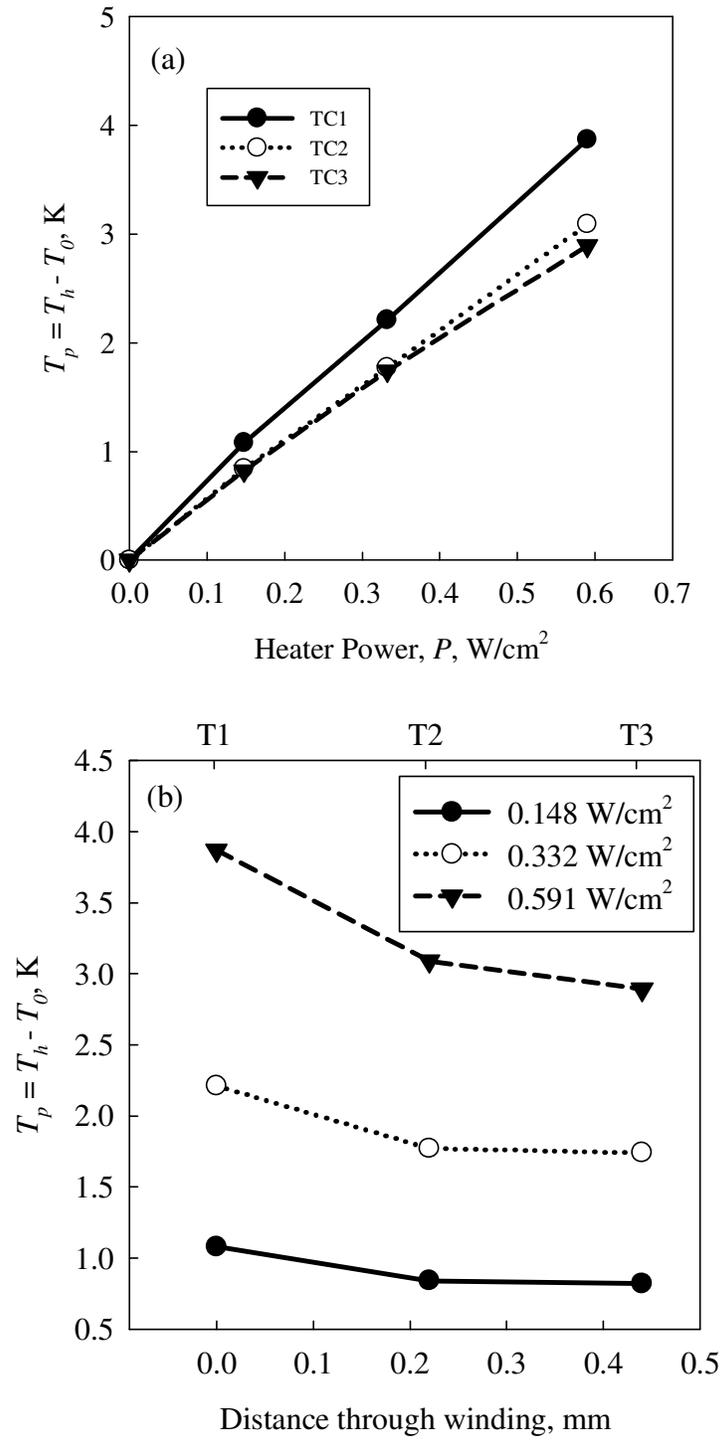

Fig. 5.

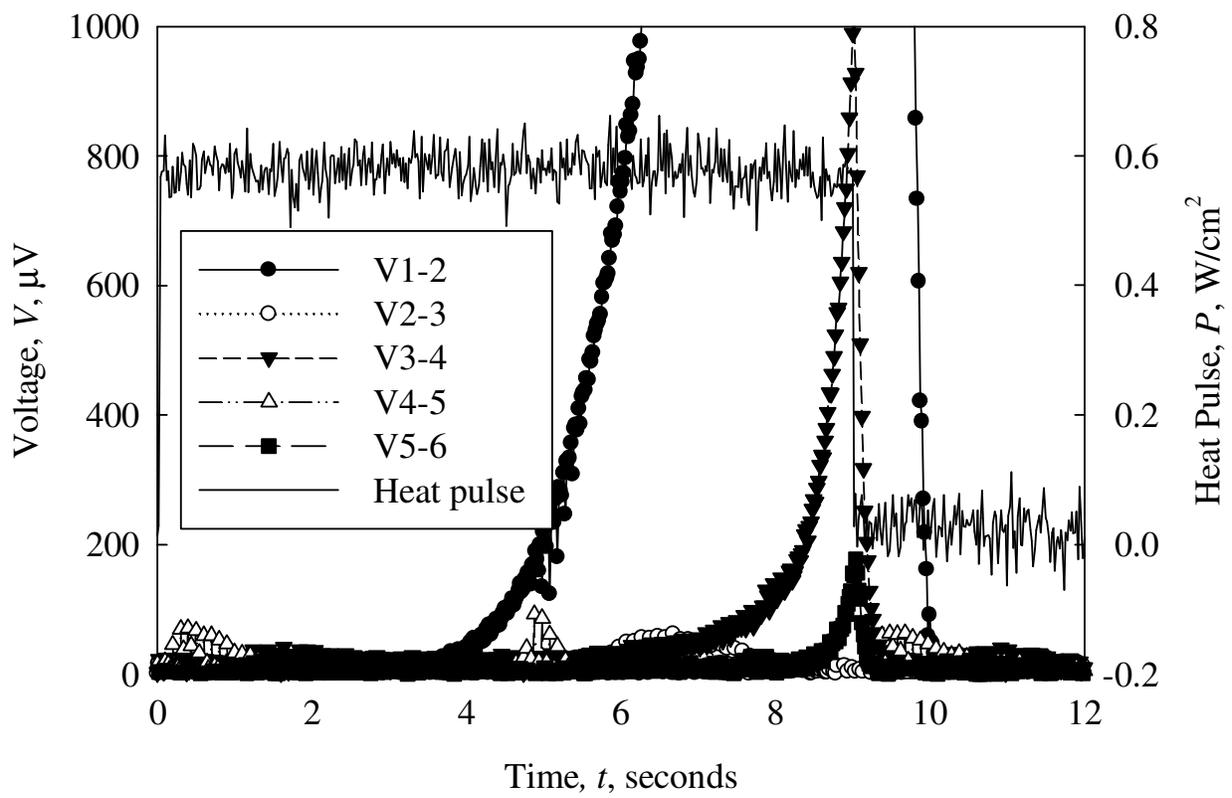

Fig. 6.

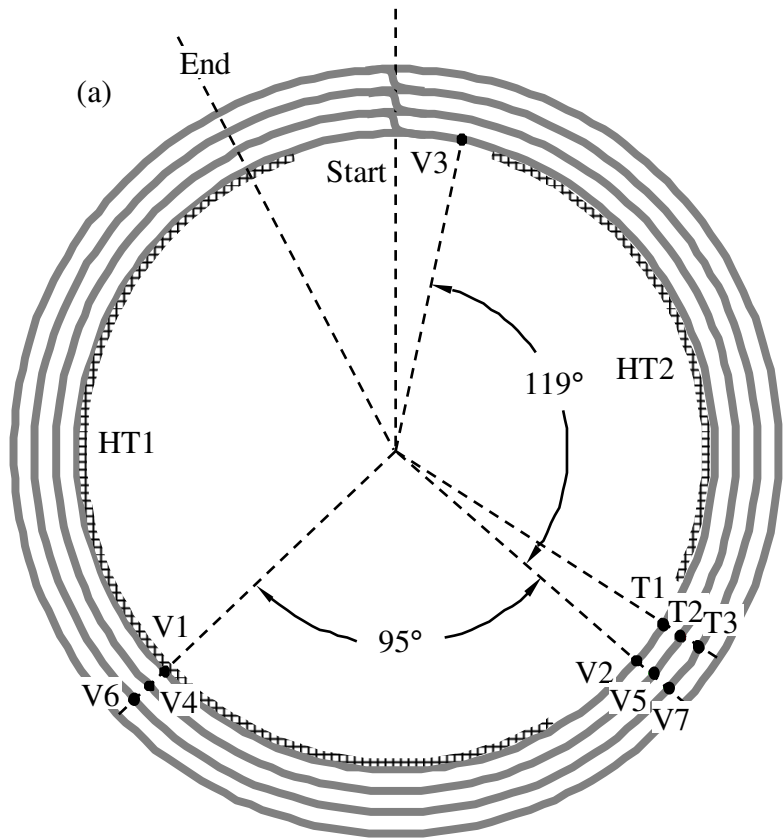

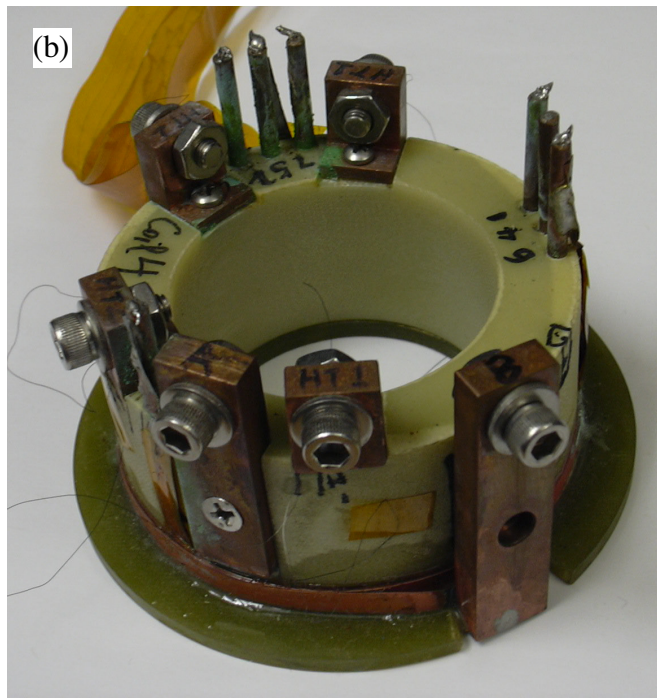

Fig. 7.

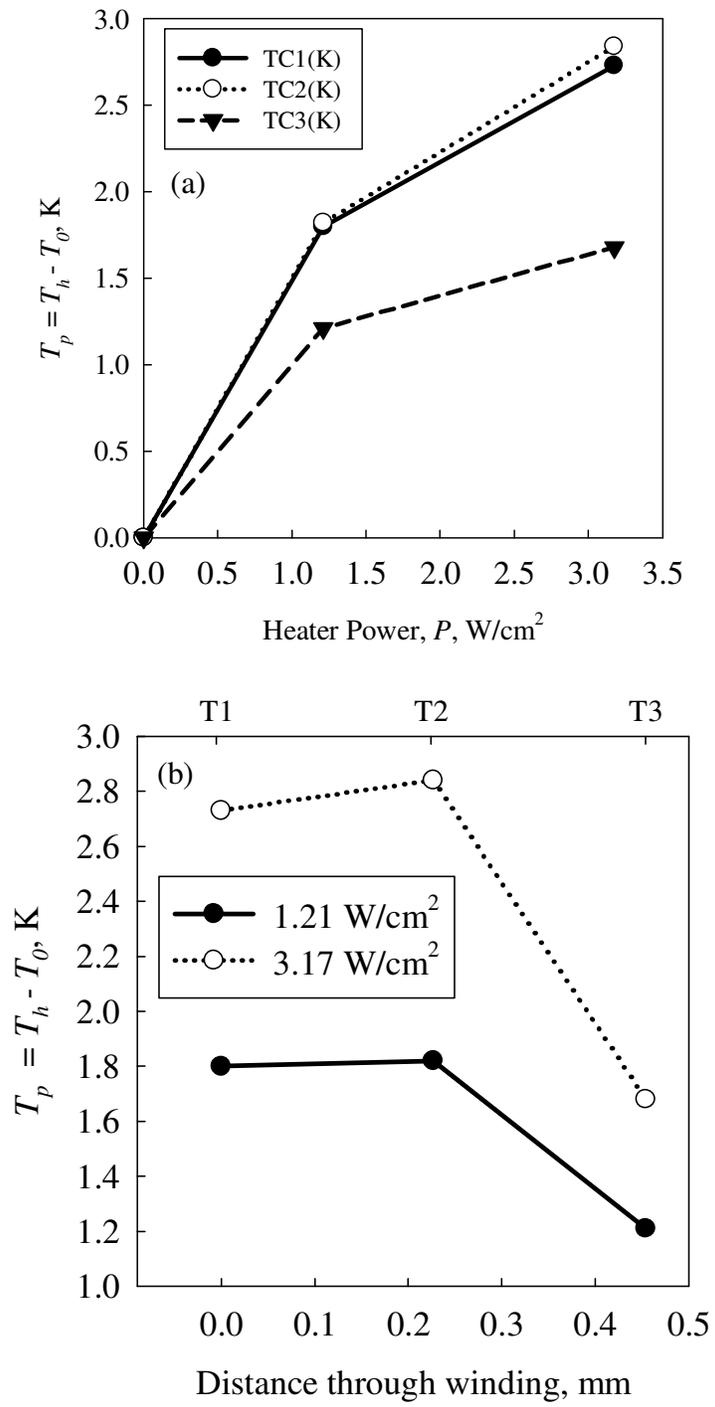

Fig. 8.

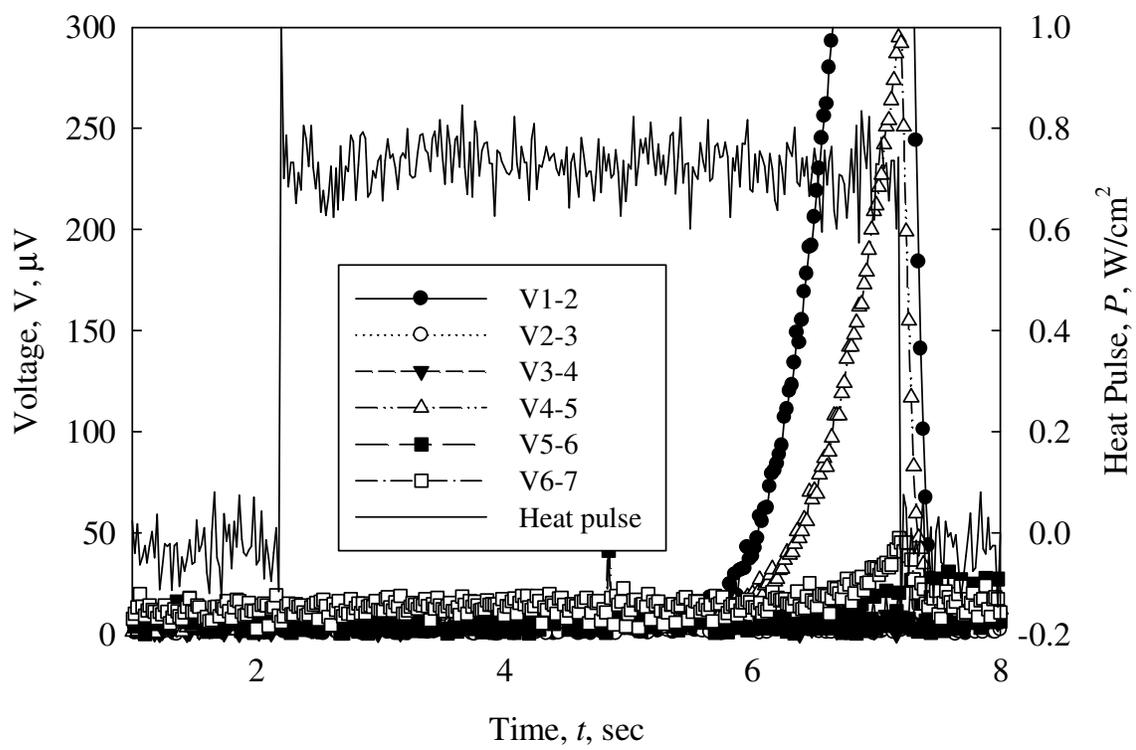

Fig. 9.

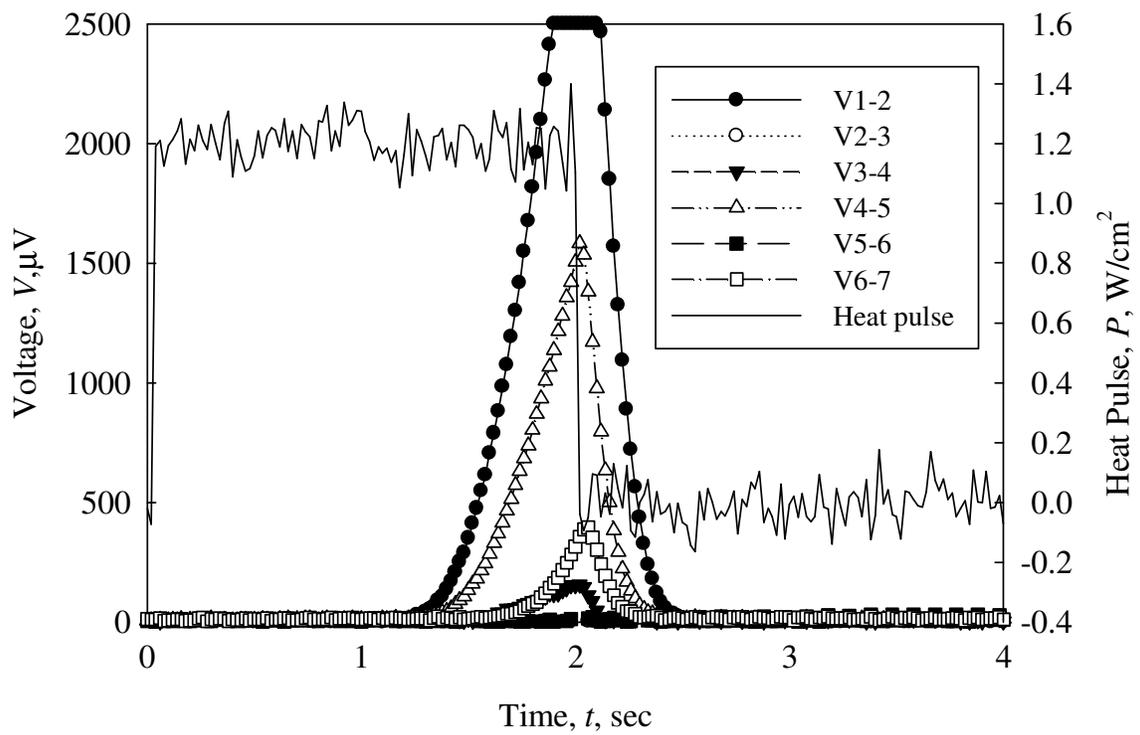

Fig. 10.

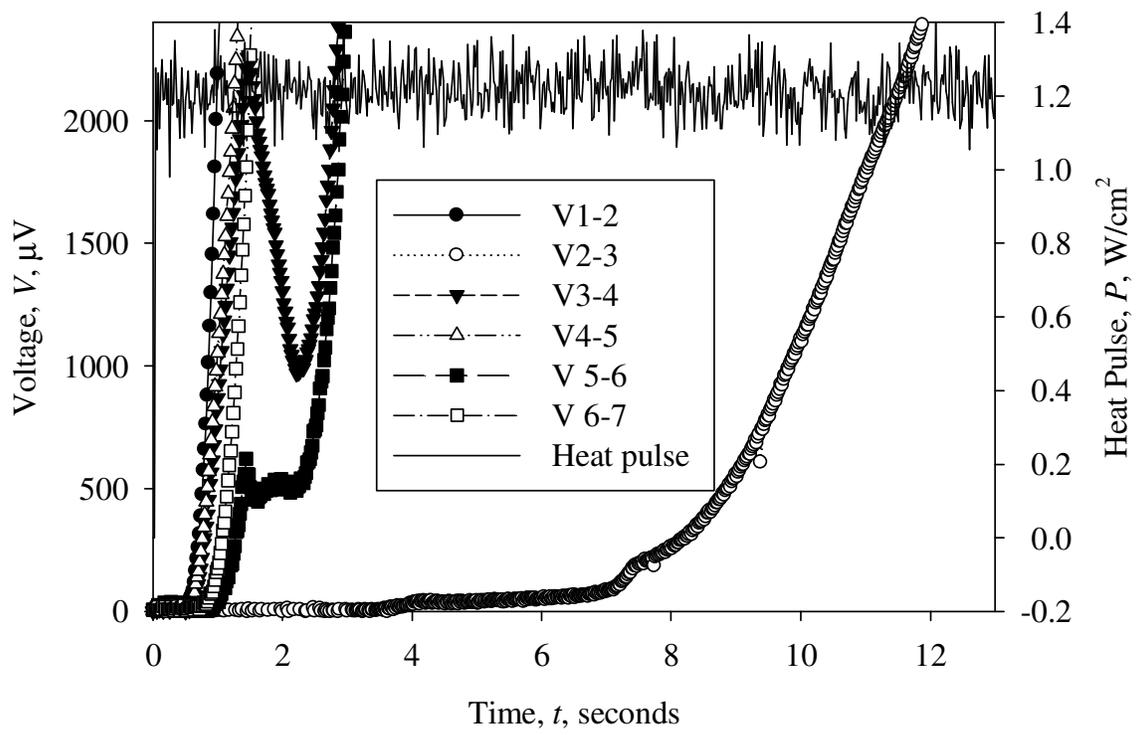

Fig. 11.

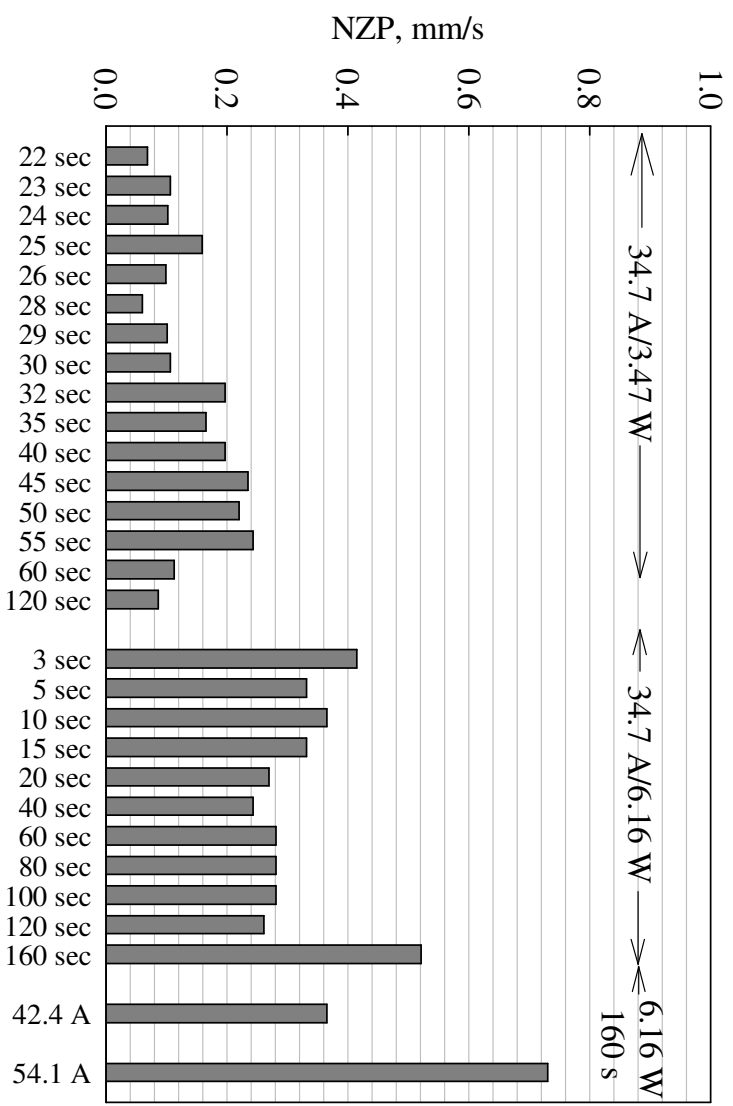

Fig. 12.

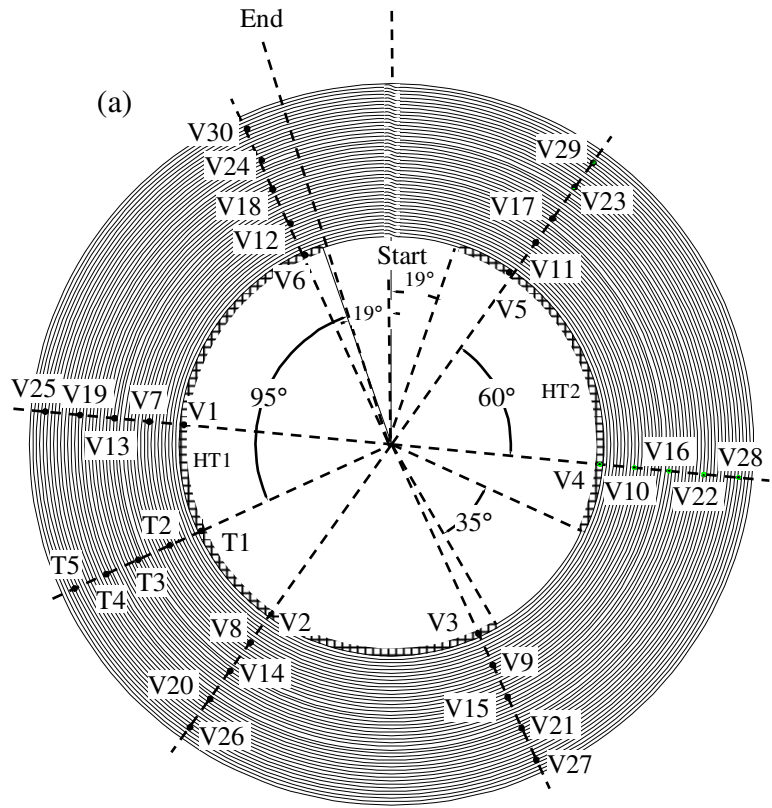

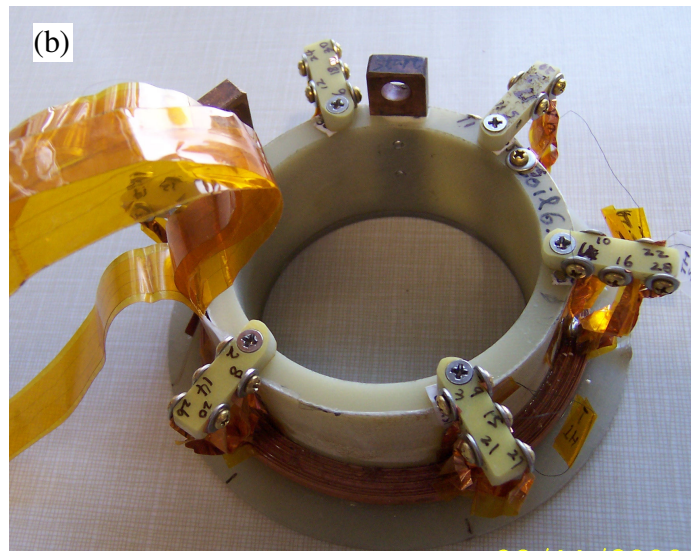

Fig.13.

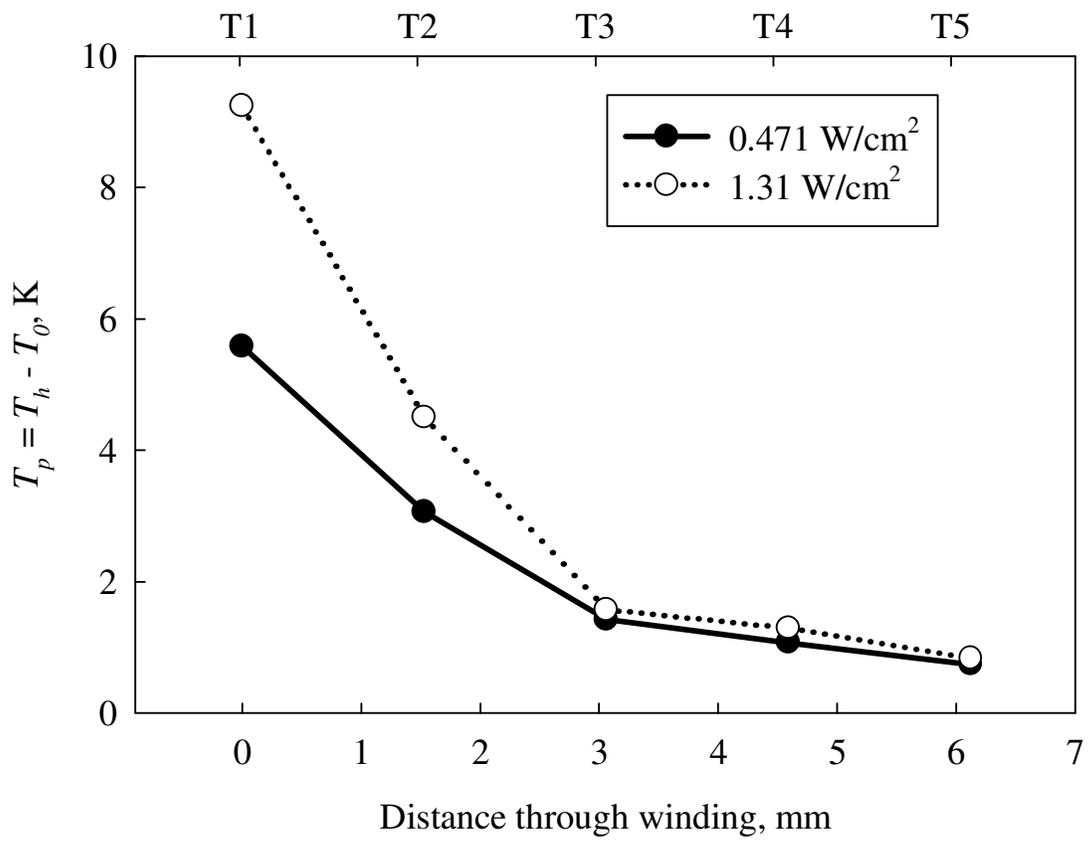

Fig. 14.

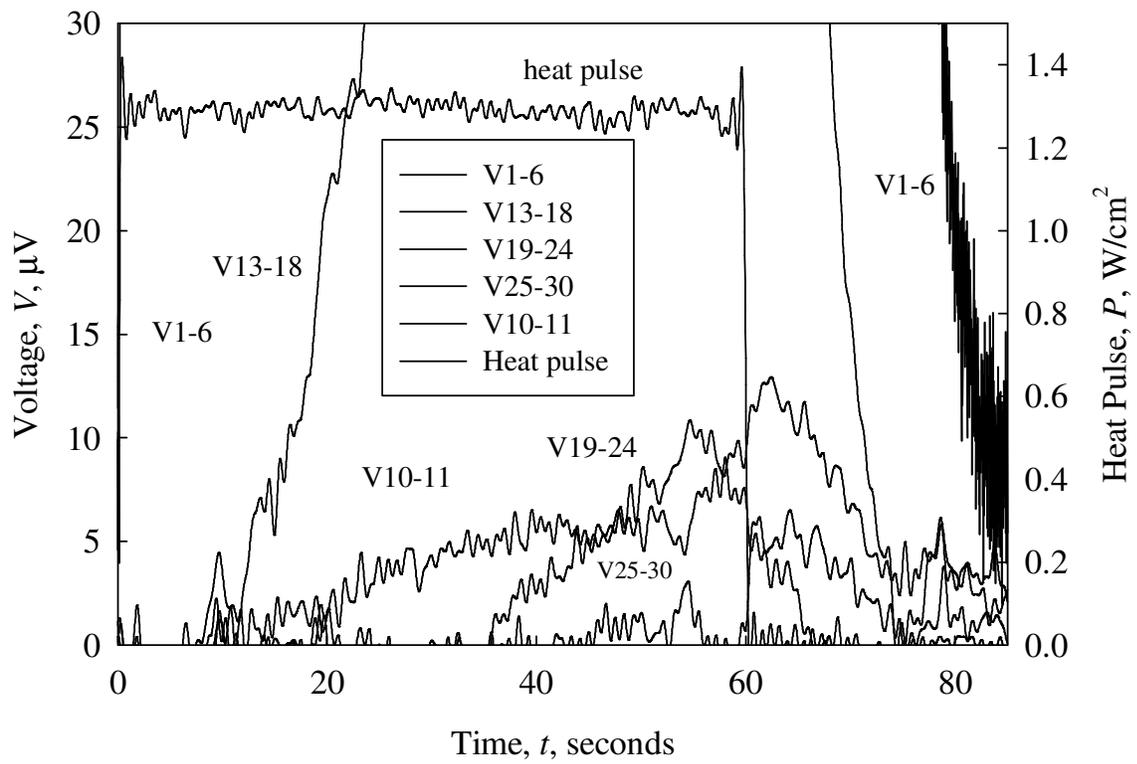

Fig. 15

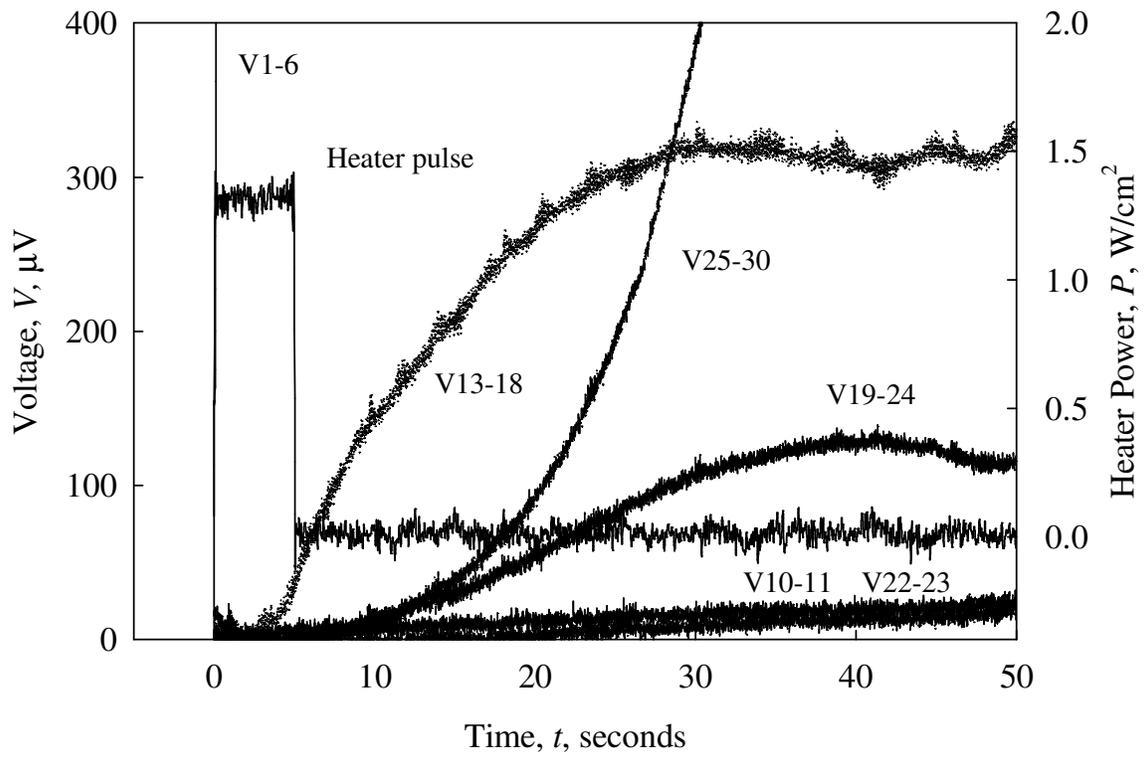

Fig.16.